# Single-photon cesium Rydberg excitation spectroscopy using 318.6-nm UV laser and room-temperature vapor cell

**JIEYING WANG,**[1,2] **JIANDONG BAI,**[1,2] **JUN HE,**[1,2,3] **AND JUNMIN WANG** [1,2,3,*]

[1] *State Key Laboratory of Quantum Optics and Quantum Optics Devices, Shanxi University, Tai Yuan 030006, Shan Xi Province, People's Republic of China*

[2] *Institute of Opto-Electronics, Shanxi University, Tai Yuan 030006, Shan Xi Province, People's Republic of China*

[3] *Collaborative Innovation Center of Extreme Optics, Shanxi University, Tai Yuan 030006, Shan Xi Province, People's Republic of China*

*\* wwjjmm@sxu.edu.cn*

**Abstract:** We demonstrate a single-photon Rydberg excitation spectroscopy of cesium (Cs) atoms in a room-temperature vapor cell. Cs atoms are excited directly from $6S_{1/2}$ ground state to $nP_{3/2}$ (n = 70 - 100) Rydberg states with a 318.6 nm ultraviolet (UV) laser, and Rydberg excitation spectra are obtained by transmission enhancement of a probe beam resonant to Cs $6S_{1/2}$, F = 4 - $6P_{3/2}$, F' = 5 transition as partial population on F = 4 ground state are transferred to Rydberg state. Analysis reveals that the observed spectra are velocity-selective spectroscopy of Rydberg state, from which the amplitude and linewidth influenced by lasers' Rabi frequency have been investigated. Fitting to energies of Cs $nP_{3/2}$ (n = 70 -100) states, the determined quantum defect is 3.56671(42). The demodulated spectra can also be employed as frequency references to stabilize the UV laser frequency to specific Cs Rydberg transition.

## 1. Introduction

Rydberg atoms with principal quantum number n>>1 have long been actively investigated. Their huge polarizabilities ($\sim n^7$) and long lifetime ($\sim n^3$) give rise to strong dipole-dipole interactions [1], long interaction time [2], and extreme sensitivity to external electric fields [3, 4]. Precision spectroscopy of such state enables applications in metrology [5, 6] and quantum information processing [7-9]. In particular, strong long-range interactions between highly excited Rydberg atoms can lead to Rydberg blockade [10, 11]. The ability to turn the interaction on and off would provide diverse applications for quantum communication [12, 13]. For alkali metal atoms, direct excitation from the ground state to the desired Rydberg states usually requires high-power ultraviolet (UV) laser which is now not commercially available; thus, people commonly prefer a two-photon excitation configuration. However, compare to single-photon Rydberg excitation, two-photon scheme have following disadvantages: decoherence due to photon scattering from the lower and upper transitions, and light shift of involved ground state and Rydberg state due to the lower and upper excitation lasers. Moreover, the selection rules of electrical dipole transition allow only excitation of nS and nD Rydberg states in two-photon scheme, and prevent access to nP Rydberg states. Up to now, experiments of single-photon Rydberg excitation are rare.

Recently, fiber lasers, fiber amplifiers, as well as efficient nonlinear frequency conversion materials have enabled development high-power UV laser for single-photon excitation of Rydberg states. In 2004, Tong *et al* [14] reported single-photon excitation of rubidium (Rb) atoms, and observed the local blockade with field ionization detection in a cold atomic ensemble. While this detection method has high efficiency and discrimination, the detected atoms are removed from the system and cannot be reused. For quantum information applications, nondestructive detection of Rydberg states is necessary. Using direct excitation, Biedermann's group [15, 16] recently demonstrated Rydberg blockade and controlling quantum entanglement between two single cold cesium (Cs) atoms trapped in two optical dipole traps with optical technique. For purely optical detection in a room-temperature vapor cell, most laser spectra of Rydberg atoms are two-photon scheme [6, 17, 18]. Therefore, few research involve nP Rydberg states, especially those with high principal quantum numbers (n>70). In 2009, Thoumany *et al* [19] reported single-photon excitation spectroscopy of Rb 63P state with a 297 nm UV laser in Rb vapor cell.

Here, we demonstrate an optical detection of a single-photon Rydberg excitation spectroscopy of Cs atoms in a room-temperature vapor cell employing the velocity-selective spectra. A 318.6 nm laser couples Cs $6S_{1/2}$, F = 4 - $nP_{3/2}$ Rydberg transition, whereas an 852.3 nm laser probes the weak excitation signal. The dependences of the spectral linewidth and amplitude on the Rabi frequencies of the coupling and probe lasers are investigated. From the Rydberg spectra, energies are obtained for the $nP_{3/2}$ states for n = 70 -100, and the quantum defect value of 3.56671(42) for $nP_{3/2}$ states is determined by fitting the experimental data. The demodulated excitation spectrum is employed to lock the UV laser frequency to a specific Cs Rydberg transition.

## 2. Experimental principle and setup

An optical detection of Rydberg excitation is different from that of the first excited state. The low transition probability and the small light-scattering cross sections lead to very weak absorption signals. To increase the Rydberg signal, an alternative detection method is adopted. The direct Rydberg transitions are detected by an enhanced transmission of an 852 nm probe beam resonant to Cs $6S_{1/2}$, F = 4 - $6P_{3/2}$, F' = 5 cycling transition. Figure 1(a) shows a V-type three-level system interacting with two laser fields composed of the ground state |c> ($6S_{1/2}$, F = 4), an excited state |b> ($6P_{3/2}$, F' = 5 or $6P_{3/2}$, F' = 4), and a Rydberg state |a> ($nP_{3/2}$). Because of the long lifetime (~100 μs) of $nP_{3/2}$ (n = 70 - 100) Rydberg states [2], some atoms populated on $6S_{1/2}$, F = 4 state will be transferred to the Rydberg state as the frequency of 318.6 nm laser is scanned across the Rydberg transition. The Rydberg excitation from the $6S_{1/2}$, F = 4 state leads to the enhanced transmission of the 852.3 nm probe laser.

For a three-level scheme, the observed spectrum in a room-temperature vapor cell is usually velocity-dependent [20, 21]. Taking into account Doppler effect, the probe laser's frequency for atoms with different velocity groups is:

$$v = v_0 \left(1 + \frac{v_z}{c}\right) = v_0 + \Delta_p \tag{1}$$

where $v_0$ is the reference frequency; $c$ is the speed of light in vacuum; $v_z$ is the atomic velocity component along the direction of the probe beam; $\Delta_p = v_0 \cdot v_z / c$ is the detuning of probe laser. Setting Cs $6S_{1/2}$, F = 4 - $6P_{3/2}$, F' = 5 cycling transition as the reference frequency, for atoms with velocity of $v_z = 0$, the detuning of the probe laser is $\Delta_p = 0$; for atoms with velocity groups of $v_z = $ -213.94 m/s, $\Delta_p$ is -251 MHz, which will result in resonance with hyperfine transition of F = 4 - F' = 4. In view of the wavelength mismatch, if the frequency of the 318.6 nm laser matches these atomic groups to $nP_{3/2}$ states, Rydberg excitation signals can be detected. The corresponding detuning of the co-propagating coupling laser $\Delta_c$ can be given as $\Delta_c = \Delta_p (\lambda_p / \lambda_c)$, where $\lambda_p$, $\lambda_c$ are the wavelengths of the probe and coupling lasers, respectively.

A schematic of the experimental setup is shown in Fig. 1(b). The high-power 318.6 nm UV coupling laser is generated by the cavity-enhanced second harmonic generation following sum-frequency generation of two infrared lasers at 1560.5 and 1076.9 nm. As discussed previously, ~2 W tunable UV laser has been achieved [22, 23]. The frequency of the coupling laser is tuned to Rydberg transition $6S_{1/2}$, F = 4 - $nP_{3/2}$ (n = 70 - 100) over the range of 318.5 - 318.7 nm. The weak probe laser is produced by an 852 nm distributed-Bragg-reflector (DBR) diode laser, and its frequency is locked to Cs $6S_{1/2}$, F = 4 - $6P_{3/2}$, F' = 5 cycling transition using polarization spectroscopy (PS). A waveguide-type electro-optic modulator (EOM) (Photline, NIR-MX800) at 852 nm is used to calibrate the frequency interval. The coupling and probe beams are collimated to diameters of ~1.6 and ~1.3 mm ($1/e^2$), respectively, and both are linearly polarized. The probe laser is split into two beams of equal power before entering a 10-cm-long fused-quartz Cs cell, and one beam is superposed with the 318.6 nm coupling beam using a dichroic mirror. The Cs vapor cell is placed in a μ-metal-shielded enclosure to reduce residual magnetic fields. Transmission of the probe beam is detected with a differential photodiode (DPD) (New Focus, 2107) after the laser passing through another dichroic mirror When the UV laser is scanned over a specific single-photon Rydberg transition, some atoms are transferred to Rydberg state, enhancing transmission of the probe beam that is overlapped with the coupling laser. The frequency of the coupling beam is calibrated by a wavelength meter (Toptica-Amstrong, HighFinesse WS-7). The detected signal is demodulated with a lock-in amplifier (SRS, SR830), creating an error signal. The error signal is fed back to the piezoelectric transducer of the 1076.9 nm fiber laser to compensate the frequency deviation using a proportion and integration amplifier (PI) (SRS, SIM960). The lock-in demodulation is enabled by UV laser frequency modulation with an acousto-optic modulator (AOM2) (Gooch & Housego, I-M110-3C10BB-3-GH27), and the modulation signal from the lock-in amplifier is added to the 110 MHz central frequency. Two AOMs are used to compensate the frequency shift from a single AOM, this ensure that the UV laser frequency is locked to a Rydberg transition.

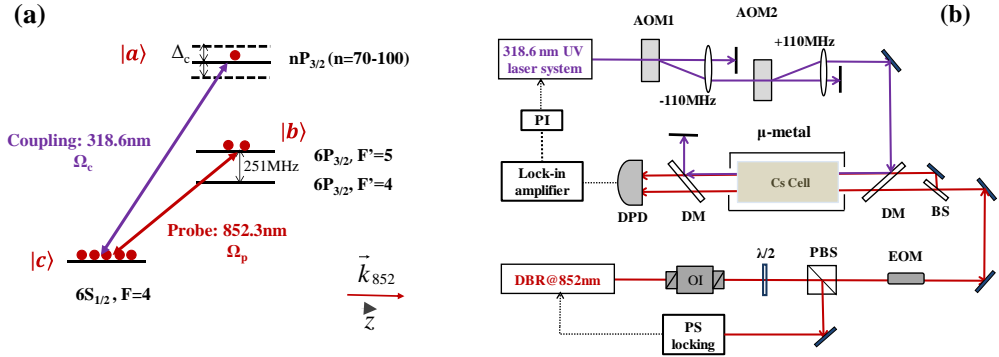

Fig. 1. (a) Relevant hyperfine levels for Cs atomic single-photon Rydberg excitation. The 852.3 nm probe laser is resonant on the transition $6S_{1/2}$, F = 4 - $6P_{3/2}$, F' = 5, and the 318.6 nm coupling laser is scanned over the Rydberg transition $6S_{1/2}$, F=4 - $nP_{3/2}$. (b) Schematic of the experimental setup. The 318.6 nm coupling laser co-propagating with the 852.3 nm probe laser in a 10-cm-long Cs vapor cell. DBR: distributed-Bragg-reflector diode laser; OI: optical isolator; λ/2: half-wave plate; PBS: polarization beam splitter cube; DM: dichroic mirror; BS: 50:50 beam splitter; AOM: acousto-optic modulator; PI: proportion and integration amplifier; EOM: electro-optic modulator; DPD: differential photodiode; PS: polarization spectroscopy.

## 3. Velocity-selective spectra of Cs $6S_{1/2}$ - $71P_{3/2}$ Rydberg excitation

Figure 2(a) presents a single-photon excitation $71P_{3/2}$ Rydberg spectrum when the probe laser is locked to $6S_{1/2}$, F = 4 - $6P_{3/2}$, F' = 5 cycling transition and coupling laser is scanned over the $6S_{1/2}$, F = 4 - $71P_{3/2}$ Rydberg transition. A small transmission peak appears when the UV laser frequency is blue-detuned to the resonance frequency of $71P_{3/2}$ state. The peak at zero detuning comes from the Rydberg excitation of the atoms with the velocity groups $v_z = 0$ and $v_z = 213.9$ m/s, and the small peak at blue detuning side is due to Cs $6S_{1/2}$, F = 4 - $6P_{3/2}$, F' = 4 hyperfine transition.

A radio-frequency modulation combined with the velocity-selective spectrum is used to calibrate the frequency interval. The spectral resolution is limited only by the signal linewidth.

Figure 2(b) is a sideband calibration result with modulation frequency of 70 MHz for 852.3 nm probe laser. In view of the wavelength mismatch of $\lambda_p / \lambda_c \approx 2.675$, the observed hyperfine interval turns into ~187 MHz. While the modulation frequency of EOM is 251 MHz, the +1-order sideband of the main peak is superposed on the small peak. Using $\Delta_c = \Delta_p (\lambda_p / \lambda_c)$, the observed hyperfine interval of the spectra becomes ~671 MHz.

To further verify that the excitation signal arise from the hyperfine splitting rather than fine splitting of nP Rydberg states as [19], the interval between the two peaks is measured using a wavelength meter as the principal quantum number n is varied from 70 to 100. Result shows that the UV detuning is a constant of ~670 MHz, and no decrease with increasing principal quantum number. Furthermore, the direct excitation oscillator strength of Cs $nP_{1/2}$ state is nearly four orders of magnitude smaller than that of Cs $nP_{3/2}$ states [15], and can't be observed here. The main transmission peak of the co-propagation scheme is used for the following experiments.

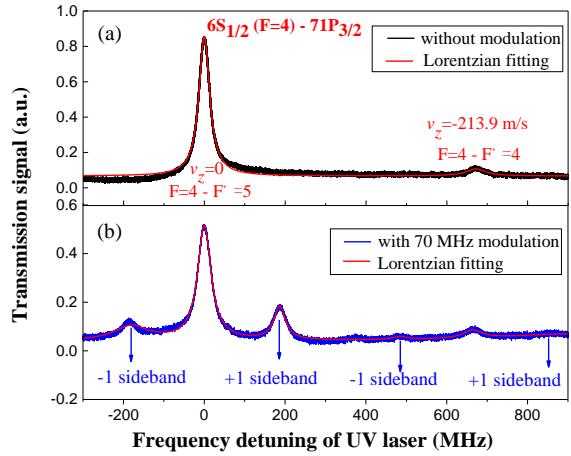

Fig. 2. (a) The excitation spectra of $6S_{1/2}$, F = 4 - $71P_{3/2}$ Rydberg transition in a Cs vapor cell when the 852.3 nm probe laser is locked to Cs $6S_{1/2}$, F = 4 - $6P_{3/2}$, F' = 5 cycling transition. The Rabi frequencies of coupling and probe beams are ~0.30 and ~8.53 MHz, respectively. (b) Sideband calibration result with a frequency modulation of 70 MHz for 852.3 nm probe laser, considering the Doppler factor of $\lambda_p / \lambda_c \approx 2.675$, the observed hyperfine interval becomes ~671 MHz. Red curve is a multi-peak Lorentzian fitting.

*3.1 Linewidths and amplitudes of Rydberg spectra versus laser intensity*

The excitation probability of single-photon Rydberg transition for higher principal quantum number n is quite small compared to that of the first excited state. For example, the oscillator strength for Cs $6S_{1/2}$ - $84P_{3/2}$ transition is $6 \times 10^{-8}$ [15], while it is 0.7164 for Cs $6S_{1/2}$ - $6P_{3/2}$ transition. Thus, it is critical for precision spectroscopy to improve the signal-to-noise ratio and suppress spectral broadening of the weak excitation signal.

We first consider a simple V-type three-level system, as shown in the level scheme. A coupling beam acts on the transition $|c\rangle \leftrightarrow |a\rangle$, with a Rabi frequency of $\Omega_c$ and a frequency detuning of $\Delta_c$. The probe beam acts on the transition $|c\rangle \leftrightarrow |b\rangle$ with a Rabi frequency of $\Omega_p$ and a frequency detuning of $\Delta_p$. The absorption of the probe laser is proportional to $Im(\rho_{bc})$, where $\rho_{bc}$ is the induced polarizability on the $|b\rangle \leftrightarrow |c\rangle$ transition. From the density-matrix equations, the steady-state value of $\rho_{bc}$ is given by [24]:

$$\rho_{bc} = i\Omega_p \frac{(\gamma_{ac}\Gamma_{ac} + 2\Omega_c^2)(i\Delta_p + \gamma_{ac})}{[(i\Delta_p + \Gamma_{bc})(i\Delta_p + \Gamma_{ab}) + \Omega_c^2](\gamma_{ac}\Gamma_{ac} + 4\Omega_c^2)} \quad (2)$$

Where $\gamma_{ij}$ and $\Gamma_{ij}$ denote the spontaneous emission rates and the decay rates from $|i\rangle$ to $|j\rangle$, respectively. $|a\rangle \leftrightarrow |b\rangle$ is a dipole forbidden transition, $\gamma_{ab} \sim 0$.

The minimum value of the probe absorption occurs at $\Delta_p = 0$, corresponding to the peak of $6S_{1/2}$, F = 4 - $6P_{3/2}$, F' = 5 transition, and the minimum value $A_{min}$ is:

$$A_{min} = \Omega_p \frac{\gamma_{ac}(\gamma_{ac}\Gamma_{ac} + 2\Omega_c^2)}{(\Gamma_{bc}\Gamma_{ab} + \Omega_c^2)(\gamma_{ac}\Gamma_{ac} + 4\Omega_c^2)} \quad (3)$$

The detected amplitude of the spectral resonance should follow linear dependence on the probe Rabi frequency while the coupling Rabi frequency is fixed.

The maximum value ($A_{max}$) of the probe absorption corresponds to $\Delta_p = \pm\sqrt{\Gamma_{bc}\Gamma_{ab} + \Omega_c^2}$. The spectral width is the frequency interval of ($A_{max} + A_{min}$)/2. In the strong probe regime of hot atoms, $\Omega_p \gg \Omega_c$, the linewidth of the spectral signal obtained from formula 2, 3, and 4 turns to [25-27]:

$$w \approx w_0 + \frac{\Omega_p^2}{\gamma_{ac} + \gamma_{bc}} \quad (4)$$

Where $w_0$ represents the broadening from other broadening mechanisms. The linewidth of the spectra will follow the square dependence on the probe Rabi frequency.

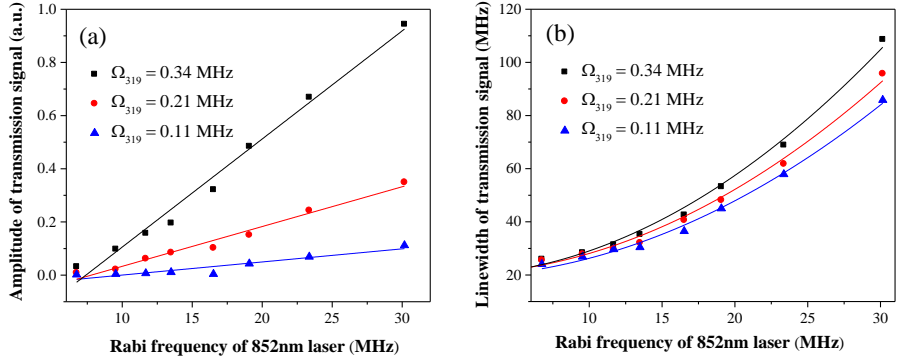

Fig. 3. Amplitudes (a) and linewidths (b) of Cs $6S_{1/2}$, F = 4 - $71P_{3/2}$ single-photon Rydberg transmission spectra as a function of probe Rabi frequency for indicated coupling Rabi frequencies. Scatters represent the experimental data, while the solid lines are the fitting results. The measured narrowest linewidth is 24 MHz for $\Omega_{319} \sim 0.11$ MHz and $\Omega_{852} \sim 6.77$ MHz.

To investigate the influence of the coupling and probe beams on the single-photon Rydberg spectra, we perform the similar measurements for various probe and coupling Rabi frequency. An EOM with a 70-MHz modulation frequency, corresponding to a UV detuning of ~187 MHz, is used to calibrate the linewidth. Figure 3 shows the measurement (scatters) of the amplitude (a) and linewidth (b) for indicated coupling Rabi frequency, together with the theoretical fitting (lines) with Eq. (3) and (4), respectively. The amplitude of the spectral peak increase significantly with both coupling and probe Rabi frequencies. As is predicted with the absorption characteristics of the probe laser, the amplitude increase linearly with probe Rabi frequency while the coupling Rabi frequency is fixed. Since the parameters $\gamma_{ij}$ and $\Gamma_{ij}$ related to Rydberg state ($|a\rangle$) in Eq. 3 are difficult to accurately determine, we regard the slope of the linear function as an overall fit parameter. The tendency for the experimental data and the fitting lines are basically coincident.

The linewidth of the main peak also follows the trend predicted by Eq. (4), as shown in Fig. 3 (b). From the fitting results, we obtained $w_0$ ~19.60 MHz, $\gamma_{ac}$ varies from 5.4 - 8.7 MHz as $\Omega_{319}$ changing from 0.11 - 0.34 MHz. In ideal condition, $\gamma_{ac}$ is a constant (<2 kHz) for Cs $71P_{3/2}$ state. However, we find that it increases by several orders of magnitude. We interpret the increase by the intricate external decoherence mechanism rather than intrinsic property of the Rydberg atoms. Besides, the deviation between the theoretical model and the experimental scheme may introduce inaccuracy. The previously measured $\gamma_{ac}$ in a cold single-atom system is about 0.2 MHz [28]. We analyze the 19.6 MHz-linewidth includes contributions from other broadening mechanisms: nature linewidth of ~14 MHz (~2.675×5.2 MHz) [19], transit-time broadening of ~450 kHz (~2.675×168 kHz), and collisional broadening [27]. When the Rabi frequency of the probe laser exceeded ~33.0 MHz, the peak begins to undergo Autler-Townes splitting. The observed narrowest linewidth is 24 MHz for $\Omega_{319}$ ~0.11 and $\Omega_{852}$ ~6.77 MHz. The factor that influence the linewidth of Rydberg spectra is very complicated. Broadening of Rydberg states in different pressure was qualitatively investigated [29], and the observed narrowest linewidth was ~10 MHz. Spectral broadening of Rydberg resonance lines resulting from high-power excitation laser in cold atoms was also observed [30]. Narrow linewidth with high signal-to-noise ratio is important for stabilizing the laser frequency to specific Rydberg transition and for precision spectroscopy.

### 3.2 Measurement of the quantum defect for Cs $nP_{3/2}$ states

Quantum defect is an important parameter for Rydberg atoms. Because Cs atoms have a big $Cs^+$ core, the quantum defects for the lower orbital angular moment states (S, P, D) are relatively large. Energy levels of Cs atoms are predicted with high precision by the quantum defect theory, where energies of the nP states can be represented by the modified Ritz formula [2]:

$$E(n,l) = E_\infty - \frac{R_{Cs}}{(n-\delta_{n,l})^2} - E_f \tag{5}$$

where $E(n,l)$ is energy of the level with a principal quantum number n and an angular quantum number $l$, $E_\infty$ = 31406.46766 cm$^{-1}$ is the ionization threshold energy, $R_{Cs}$ is the Cs Rydberg constant of 109736.86274 cm$^{-1}$, $E_f$ = 4.0217764 GHz is the hyperfine frequency shift of the Cs $6S_{1/2}$, F = 4 hyperfine component from the $6S_{1/2}$ ground state. The parameter $\delta_{n,l}$ is the quantum defect given as follows [2]:

$$\delta_{n,l} = \delta_0 + \frac{\delta_1}{(n-\delta_0)^2} + \frac{\delta_2}{(n-\delta_0)^4} + \ldots \tag{6}$$

For $n \geq 70$, the quantum defect can be treated as a constant for the same angular momentum states as the higher order terms of $\delta_{n,l}$ are small and can be neglected.

The measured energy values (dots) of the Cs $nP_{3/2}$ states (n = 70 - 100) are plotted in Fig. 4, where the measurement range of principal quantum number n is mainly limited by the tuning range of our UV laser system. Fitting to the measurement data using Eq. 5 (solid line) yields a quantum defect value of 3.56671(42). The tiny difference with the previous value of 3.55925 measured with n = 9 - 50) [30] is that our experimental scheme have no external electric-field control part, which will introduce the shift of the nP levels. Besides, the accuracy is also limited by the wavelength meter and the linewidth of the 852 nm DBR laser.

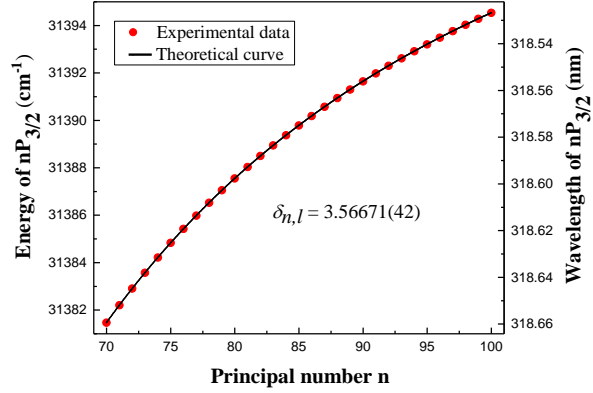

Fig. 4. Energies of Cs $nP_{3/2}$ Rydberg states (n = 70 - 100). The dots are experimental data, and the solid line is the fitting curve. The quantum defect value of Cs $nP_{3/2}$ states is 3.56671(42) is obtained, where the error is statistical error.

## 4. UV laser frequency stabilization with single-photon Cs Rydberg spectra

After optimizing the linewidth and the signal-to-noise ratio, we choose Rydberg spectrum of Cs $6S_{1/2}$, F = 4 - $71P_{3/2}$ transition with a linewidth of 27.3 MHz ($\Omega_{319}$ ~ 0.11 MHz and $\Omega_{852}$ ~ 9.50 MHz) as the reference frequency to stabilize the UV laser, as shown in the upper part of Fig. 5(a). Here the probe laser is frequency modulated at 17 kHz by AOM2. A dispersion-like signal is obtained with the lock-in amplifier as the lower part of Fig. 5(a), which is fed back to the piezoelectric transducer of the 1076.9 nm fiber laser to compensate the frequency fluctuation using a PI controller. The estimated bandwidth of the servo loop is ~10 kHz, which is limited by the integration time (300 μs) of the lock-in amplifier. Using the main peak of the Rydberg spectra as the reference frequency, the UV laser frequency is locked to the zero-crossing point.

The Allan deviation of a locking laser system should be measured by the optical frequency comb technique or the beat note of two identical laser systems, which can accurately reveal the laser frequency stability. Unfortunately, the commercial UV optical frequency combs are very scarce and we don't have another same UV laser system. Assuming the single-photon Rydberg excitation spectrum is absolutely stable, and the reference point is not time-varying, the relative Allan deviation can be used to estimate the frequency stability. Figure 5(b) shows relative Allan deviation of the UV laser, and the relative frequency stability is estimated from the slope of the zero-crossing point in the error signal. Typical relative frequency stability is ~ $3.2 \times 10^{-10}$ (~300 kHz) for the interrogation time of 1 s, and it is ~ $1.3 \times 10^{-11}$ (~12 kHz) for the interrogation time of 56 s. It should be emphasized that the Allan deviation is originated from the error signal, thus, it only represents a lower limit of the frequency instability. However, 300 kHz is still a considerable improvement over the free-running large frequency fluctuation, which is about several megahertz over 1 s. Hence, the frequency locking technique is effective to compensate the long-term frequency drift of the UV laser. The frequency stability is sufficient to perform the single-photon Rydberg excitation of Cs atoms from $6S_{1/2}$ to nP (n = 70~100) state [15].

Another optional technique to stabilize the UV laser system is locking the 1560.5 and 1076.9 nm lasers to an ultra-stable cavity in vacuum. Cavity-locking method strongly depends on the environmental factors such as temperature and pressure while it has narrower resonance linewidth. Compared with the former locking technique, frequency stabilization with single-photon Rydberg spectra provides a more direct frequency standard. It can directly realize the laser frequency stabilization to a special Rydberg transition.

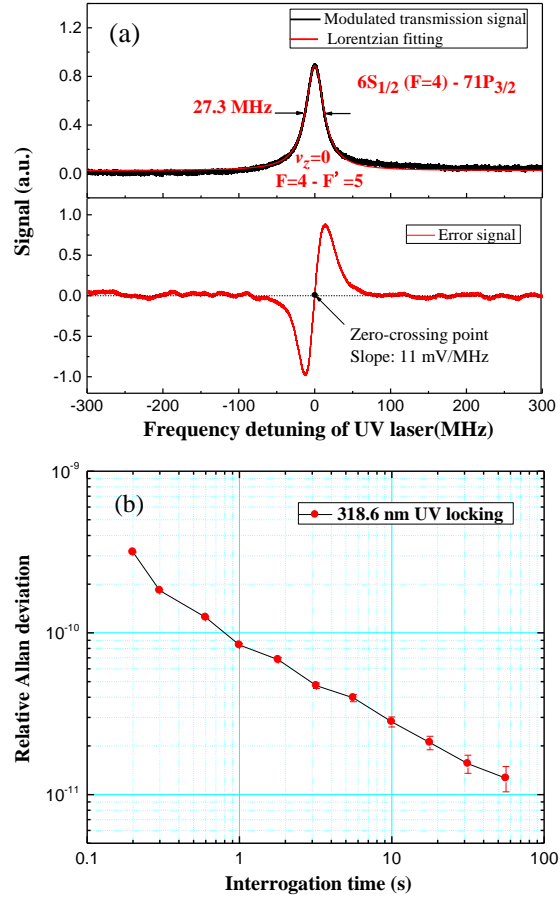

Fig. 5. (a) The frequency-modulated Rydberg spectra with $\Omega_{319} \sim 0.11$ MHz and $\Omega_{852} \sim 9.50$ MHz, and the corresponding error signal from the lock-in amplifier with a modulation frequency and an integration time of 17 kHz and 300 μs, respectively. The red curve in the upper part is a multi-peak Lorentzian fitting. (b) Relative Allan standard deviation plots show the relative frequency instability of the 318.6 nm UV laser (dots).

## 5. Summary

We have performed optical detection of a single-photon Rydberg excitation spectroscopy of Cs atoms in a room-temperature vapor cell. The velocity-selective spectra of Cs $6S_{1/2}$, F = 4 - $nP_{3/2}$ (n = 70 - 100) transitions are consistent with theoretical analysis of the Doppler mismatch. The dependences of the spectral linewidths and amplitudes on the coupling and probe beams' Rabi frequency are investigated. The measurements of Rydberg energy of Cs $nP_{3/2}$ (n = 70 - 100) states show good agreement with the theoretical fitting, which yields the quantum defect value $\delta_{nl}$ (l=1) = 3.56671(42). The measurement accuracy is mainly limited by the wavelengthmeter's accuracy and the linewidth of the 852 nm probe laser used in our experiment. After optimization of the spectral parameters, the error signal is used to lock the UV laser frequency to a specific Rydberg transition. The experimental results reveals the ability to stabilize the UV laser frequency to one of Cs $6S_{1/2}$ - $nP_{3/2}$ (n = 70 - 100) Rydberg transitions. Investigation of the high precision single-photon Rydberg spectroscopy in room-temperature Cs atoms should impact and improve the excitation and detection techniques, especially for studying nP states' energy structure of Rydberg atoms and the interaction between Rydberg atoms.

**Funding**

This work is supported by the National Natural Science Foundation of China (NSFC Grant Nos. 61475091 and 61227902), and by the National Key Research and Development Program of China (2017YFA0304502).